\documentclass[%
 aip,
 jmp,%
 amsmath,amssymb,
 reprint,%
nofootinbib]{revtex4-1}

\usepackage{graphicx}
\usepackage{dcolumn}
\usepackage{bm}
\usepackage{subfigure}
\usepackage{color} 

\begin{document}

\preprint{AIP/123-QED}

\title[]{Does the schock wave in a highly ionized non-isothermal plasma really exist  ? }

\author{A. A. Rukhadze}
\email{rukh@fpl.gpi.ru}
\affiliation{Prokhorov General Physics Institute,
Russian Academy of Sciences, Vavilov Str. 38., Moscow, 119991, Russia}
\author{S. P. Sadykova}%
\altaffiliation{Electronic mail: \textcolor{blue}{Corresponding author - s.sadykova@fz-juelich.de}}
\affiliation{Forschungszentrum Julich, J{\"u}lich Supercomputing Center, J{\"u}lich, Germany
}%
\author{T. G. Samkharadze}
\affiliation{Moscow State University of Instrument Engineering 
and Computer Science
Moscow, 107996, Stromynka str. 20,  Russia}


\date{\today}

\begin{abstract}
Here we study the structure of a highly ionizing shock wave in a gas of high atmospheric pressure. 
We take into account the gas ionization when the gas temperature reaches few orders of an ionization potential. It is shown that after gasdynamic temperature-raising shock and formation of a highly-ionized nonisothermal plasma $T_e>>T_i$ only the solitary ion-sound wave (soliton) can  propagate in this plasma. In such a wave the charge separation occurs: electrons and ions form the double electric layer with the electric field. The shock wave form, its amplitude and front width are obtained.

\end{abstract}

\maketitle

\section{\label{sec:Int}Introduction}
The shock wave  in a medium is a propagating disturbance. 
It is characterized by an abrupt, nearly discontinuous change (shock) in thermodynamic parameters of the medium (for simplicity, we consider only the ideal gas). In a case of the plane shock wave the gas parameters (pressure, temperature, density) before and after the shock are constant and homogeneous and for a strong wave they can be interrelated like the following \cite{1} :
\begin{equation}
\label{1}
\frac{P_2}{P_1}=\frac{2\gamma}{2\gamma+1}M^2, \; \frac{T_2}{T_1}=\frac{2\gamma(\gamma-1)}{(\gamma+1)^2}M^2,\; \frac{n_2}{n_1}=\frac{(\gamma+1)}{(\gamma-1)^2},
\end{equation}
 
\indent here the subscripts $1$ and $2$ are referred respectively to the gas state before and after the shock front, $M$ - Mach number is the ratio of speed of shock wave front and the speed of sound in a gas before the front, $\gamma=\frac{C_p}{C_v} $ is the ratio of  the heat capacity at a constant pressure and  the heat capacity at a constant volume (for one atomic gas $\gamma=5/3$).
These states are related to the non-isothermal shock. In a case of the isothermal shock when the thermal conductivity (for ex. thermal radiation) in a medium  is quite high the relation \ref{1}  will transform  into the following \cite{1}:
\begin{equation}
\label{2}
\frac{P_2}{P_1}=\frac{n_2}{n_1}=\frac{(3-\gamma)}{\gamma+1},
\end{equation}
From this it follows that at $\gamma=1$ the right hand-side will become unity, hence in such a  medium the shock wave does not exist. However, it does not mean that nonlinear waves can not exist in such a medium. For example (we will demonstrate it further down), in such a medium the propagation of a soliton is possible whereas the medium state before the wave front side and behind the back side of the front does not change. \\
\indent Let us return to the shock waves and discuss the structure of the transition layer in a gas. There are a lot of works devoted to such a problem mainly related to the weak shocks ($M\simeq 1$). The waves with the arbitrary intensity were considered  by I. Y. Tamm in 1947. However the work was first published in 1965 \cite{2}. In a case of the weak shock wave ($M \approx 1$) the structure of the transition layer (wave front width) defined in a framework of hydrodynamics is given by the following formula:
	\begin{equation}
\label{3}
d=\frac{P_2+P_1}{P_2-P_1}\delta,\;\; \delta=1.28 l,
\end{equation}
where $l$ - mean free path of gas particles. From Eq. (\ref{3}) follows that the front width is greater than the free path and with an increase of $P_2/P_1$ - $d\to\delta$ limits to the free path.\\
\indent In a case of the strong shock wave ($P_2/P_1>>1$) the wave front width obtained in \cite{2}  is the following:
\begin{equation}
\label{4}
d=0.503 l_1 = 2.012 l_2,
\end{equation}
where $l_1$ and $l_2$ are the particles free paths in the front side and back side of the shock wave respectively.  The results obtained in the work \cite{2} in the hydrodynamic approximation  were found in a good agreement with those obtained in a frame of the Boltzmann approximation, i.e. in the kinetic approximation. However, the simplified approximation was used: Boltzmann equation with an account of the  elastic collisions in a framework of the hard spheres.

	\section{Strong ionizing shock wave}
	
	In gases of high pressure (atmospheric and higher one) the structure of the strong shock waves described in \cite{2} is too simplified. At Mach number greater than 10 due to the gas  heating the gas ionization and generation of highly ionized plasma must occur what was not taken into consideration in \cite{2}. As a result, the structure of the strong ionizing wave must have few peculiarities. Namely, right behind the front side of the shock wave at a free path length the gas  temperature gets increased to several electron-volts leading to the gas ionization and formation of a strongly ionized non-isothermal plasma. The electron temperature in such a plasma is of an order  of a gas atom ionization potential whereas the ion temperature is much less that that of electrons. At the same time according to the gas-phase approximation Debye electron radius is less than electron free path. This phase transition occurs at an electron constant temperature of an order of an atom ionization potential. It lasts for few orders of an inverse ionization frequencies  until the whole gas gets ionized. The width of this domain amounts to few ionization free paths. As a result the fully ionized non-isothermal plasma with the electron temperature of an order of few atom ionization potentials gets formed behind the front in which the shock wave should travel.\\
	\indent We should ask ourselves: ``Is this possible ?''. - `` No , it is not!'' The point is that the acoustic oscillation mode in non-isothermal highly ionized plasma exists only when $T_e>>T_i$ - the ion sound is an isothermal sound with  $\gamma=1$ ($T_e$=const due to the high electron thermal conductivity) \cite{3,4}. According to Eq. (\ref{2}) the isothermal shock in a neutral gas at $\gamma=1$ is not possible. As it will be demonstrated below, in non-isothermal plasma when $T_e>>T_i$ only the propagation of a solitary wave (soliton) with the half-width equal to the Debye electron radius is possible. This wave represents the double electric  layer in which the electrons (electron layer) overtake the ions (ion layer) and pull them on.  The electric potential of the double layer  can be described  as a weakly nonideal wave by the following equation \cite{4}:
	\begin{equation}
\label{5}
\frac{T_e r_{De}^2}{2e}\frac{d^2\Phi}{d\xi^2}- \frac{T_e}{e}(1-\frac{V_s}{u})\Phi+\frac{1}{2}\Phi^2=0,
\end{equation}
here $\xi=x-ut$, $r_{De}=\sqrt{T_e/4\pi e^2n_0}$- Debye electron radius, where $n_e=n_0 \cdot exp(e\Phi/T_e)$, $V_s=\sqrt {T_e/M}$ - ion sound velocity, $u$ - to be determined velocity of a soliton equal to the velocity of the shock wave. \\
\indent Eq. (\ref{5}) known as the Korteweg–de Vries equation (KdV Eq.  for short) has the following explicit solution:
\begin{equation}
\label{6}
\Phi=\frac{\Phi_{max}}{ch^2 (\xi/\Delta)},
\end{equation}
where
\begin{equation}
\label{7}
1-\frac{V_s}{u}=\frac{e\Phi_{max}}{\pi T_e}<<1, \; \; \frac{e\Phi_{max}}{6 T_e}=\frac{r_{De}^2}{\Delta^2}<< 1.
\end{equation}
As one can see $u\succeq V_s$, i.e. the velocity of a solitary wave is almost equal to that of an ion sound, the wave full width at half maximum $\Delta > r_{De}$ and decreases with increasing wave amplitude.

\section{Discussion of the results and quantitative estimations}

Eqs. (\ref{6}) and (\ref{7}) are exact on implementation domain and describe exactly the solitary wave (\ref{6}) in a plasma of the highly ionizing shock wave with temperature $T_e\sim 3 I_i\sim 1\cdot 10^5$ K, where $I_i\sim 10$ eV - gas atom ionization potential. At a atmospheric gas pressure - $r_{De}\sim 7 \cdot 10^{-7}$ cm and the mean distance between plasma electrons - $r_{a}\sim 7 \cdot 10^{-7}$ cm, i.e. the gas-phase approximation condition ($r_{a}^2<< r_{De}^2$) is satisfied. According to  Eq. (\ref{7}) potential amplitude increases with decreasing $\Delta$ and at $\Delta=3 r_{De}$ (the minimum magnitude at which the equations (\ref{5}) - (\ref{7}) are valid) becomes equal to:
\begin{equation}
\label{8}
E_{max}\leq 10^7  V/cm,
\end{equation}
There exists another constraint for the implementation domain of Eq. (\ref{6}), namely, the small   plasma disturbance by a soliton field when
\begin{equation}
\label{9}
E\leq \sqrt{4\pi n_e T_e} \sim 10^8  V/cm.
\end{equation}
This field is one order higher than the field (\ref{8}). The field $E=10^7$ V/cm is high enough to create the electric discharge in a solid body (for example in a glass). When the condition (\ref{8}) is violated then the plasma gets exploded, electrons will break apart of ions and get accelerated by the field. However, this stage is out of our consideration and requires  a more detailed consideration. \\
\indent For realization of described above process in atmosphere at normal conditions the Mach number of the shock wave must be of order $M\sim \sqrt{T_e/T_0}\sim 33$. When the Mach number is quite high then the plasma explosion occurs forming the electron gas where the field will exceed the magnitude (\ref{8}) and the front width will increase with increasing field. This stage is out of our consideration here in this work. \\
In this way the strong ionizing shock wave in a gas must have a very complex structure: right in  the back side of the front at a length of the mean free path of gas particles before the front, the temperature increases to few electron volts and the gas ionization occurs. In the formed plasma   behind this temperature-raising shock traveling only of the ion-sound soliton is possible. The  soliton half-width is of order of hot electrons Debye radius. In a case of the strong shock wave the ion-sound soliton gets transformed into the extended double layer with the following electric field \cite{5}:
\begin{equation}
\label{10}
E \sim \sqrt{4\pi n_e T_e} =\sqrt{4\pi n_e T_0 T_e/ T_0}=M \sqrt{4\pi n_e T_0},
\end{equation}
where $T_0$ -gas temperature in the front-side of the front, $M>>1$ - Mach number, $n_e$ - density of a plasma  formed behind the front which is close to the neutral gas density. Relation (\ref{1}) shows that compared to the normal gas where the pressure behind the front of the strong shock wave is $M^2$ higher than that in front-side, in a plasma, the plasma gets polarized behind the front of the shock wave producing the field with the same pressure.\\
\indent According to the described above,  in non-isothermal plasma formation and traveling only of the ion-sound soliton (and not of the shock wave) is possible. From Eqs. (\ref{5}) and (\ref{6}) it follows that the soliton represents the blob of a electromagnetic field revealing not only  the  resonance formation of the soliton but also that the soliton is a resonator itself having the electromagnetic field which can lead to the electric discharge at soliton disturbance for ex. at a collision with the hard target like glass. The latter, discharges inside the window glasses, was revealed after the Chelyabinsk meteor explosion that entered Earth's atmosphere over Russia on 15 February 2013 at about 09:20 YEKT. \cite{Chel}\\
\indent It is particularly interesting to discuss the concept of the shock waves in plasmas which got the increased attention in the literature \cite{7}-\cite{10}. The short overview of these works is given in monography \cite{11}. The authors discuss plasma and the long-range Coulomb  interactions of charged plasma particles. However, they consider plasma as an one-atomic gas with the adiabatic constant $\gamma=5/3$ and the ion temperature $T_i\geq T_e$. At the same time, it is well known that the Coulomb system of charged particles can be treated as a plasma only when the Langmuir plasma frequency is much greater than the electron collisions: 
\begin{equation}
\label{11}
 \omega_{Le}^2 >>\nu_e^2.
\end{equation}
We would like to note that the sound in a plasma is isothermal (hence $\gamma=1$) and can exist in non-isothermal plasma only with  $T_e>> T_i$. Exactly in such plasma the density and temperature shocks are impossible and only the solitary waves can there exist.\\
\indent If the inverse constraint (\ref{11}) is satisfied then the peculiar plasma properties can not be revealed; plasma becomes similar to the gas of neutral particles where also the charged particles are present which play no principle role. In our point of view the authors of works \cite{7}-\cite{10} meant such plasma and the words (here citation): `` In a plasma an electron is strongly bound to the ion by the charge separation field and they move together as a whole unit'' have nothing to do with the plasma.

\begin{acknowledgments}
 A. Rukhdaze expresses his gratitude to Sergeev who has read this work in written and made many important remarks. S.P. Sadykova would like to express her gratitude to her father P. S. Sadykov  for being all the way the great moral support for her and the Helmholtz foundanion  for the financial support .
\end{acknowledgments}

\appendix

\nocite{*}

\end{document}